\documentclass[twocolumn,amsmath,amssymb,showpacs,floatfix,amsfonts]{revtex4}
\usepackage{graphicx}
\def \beq {\begin{equation}}
\def \eeq {\end{equation}}

\begin{document}
\title{Quantum Zeno effect in atomic spin-exchange collisions}
\author{I. K. Kominis}
\email{ikominis@iesl.forth.gr}

\affiliation{Department of Physics, University of Crete, Heraklion
71103, Greece} \affiliation{Institute of Electronic Structure and
Laser, Foundation for Research and Technology, Heraklion 71110,
Greece}

\date{\today}
\begin{abstract}
The suppression of spin-exchange relaxation in dense alkali-metal vapors discovered in 1973 and governing modern atomic magnetometers is
here reformulated in terms of quantum measurement theory and the quantum Zeno effect. This provides a new perspective of understanding decoherence in
spin-polarized atomic vapors.
\end{abstract}
\maketitle
\section{Introduction}
Spin-exchange collisions \cite{happer1}, brought about by the Pauli exchange interaction,
play a dominant role in the physics of spin-polarized atomic vapors and their applications \cite{happer_rmp}. Spin-exchange collisions
are responsible not only for the very useful transfer of spin-polarization from one atomic species to another \cite{spin_ex_prl},
but also for the detrimental effect they have on spin coherence, i.e. spin-exchange collisions cause decoherence \cite{appelt}.
The spin-coherence lifetime poses fundamental limitations to precision measurements involving spin-polarized atoms \cite{kom_prl}, as for example
measurements of a small magnetic field (or a small Larmor frequency) performed with atomic magnetometers \cite{budker,romalis_prl,nat_mag,mabuchi}.
However, it was early on realized \cite{happer_tang,happer_tam} that decoherence due to spin-exchange collisions can be suppressed if the
spin-exchange rate is large enough relative to the frequency scale set by the atomic Larmor precession in an external magnetic field.
In this work we will re-interpret this result in terms of quantum measurement theory \cite{braginsky}. In particular, we will reformulate this
in terms of the quantum Zeno effect \cite{misra}, the essence of which is that a frequent enough interrogation of a quantum system fundamentally alters its time evolution. We will also consider the physical information on the atomic spin state provided by these collision-induced measurements. From this perspective, we will also describe another kind of spin-dependent atomic collisions, namely spin-destruction collisions.
The latter also lead to decoherence, which however is monotonically increasing with the collision rate, contrary to spin-exchange collisions. Whereas both kinds of collisions can be understood as performing a quantum measurement of the atomic spin coherence, they fundamentally differ on the route taken by the information provided by these measurements. In spin-exchange collisions, some information is in principle available, whereas in spin-destruction collisions the information is irretrievably lost in the environment. The reason for elaborating on this
alternative perspective on spin-exchange collisions is that it motivated a recently discovered analogy to
a seemingly different physical system, namely the charge-recombination of radical-ion-pairs \cite{kom1}.
\section{Spin Exchange Collisions as an information-rich Quantum Measurement}
In describing quantum measurements, we usually distinguish between
the quantum system under consideration and the quantum probe which
is an auxiliary quantum system. The probe interacts with the
quantum system, all information on which is later extracted by
performing measurements on the quantum probe \cite{braginsky}. The dissipative
interaction of an open quantum system with its environment can
also be molded into the previous picture, only now the probe
system describing the environmental degrees of freedom is
unobserved, i.e. information about the quantum system
irretrievable leaks into the environment. While decoherence is present in both cases, in the former it is
due to the unavoidable back-action of the probe onto the system, whereas in the latter due to information leakage to the
environment.

In the specific case of $N$
alkali-metal atoms confined in a cell, each atom is the quantum
system, whereas all other atoms form a multitude of quantum
probes. This distinction obviously fades away as we describe the
combined system of $N$ atoms, the behavior of which is an average
over $N$ separate quantum systems. The system degrees of freedom
are embodied in the atomic spin state, described as usual \cite{appelt} by
the atom's $2(2I+1)$-dimensional ground state Hilbert space, where $I$ is the atom's nuclear spin. The
environmental degrees of freedom are found in the practically classical translational
angular momentum of the atoms. The binary spin-exchange interaction
Hamiltonian of two colliding atoms with electron spin $\mathbf
s_{1}$ and $\mathbf s_{2}$ is of the form $h_{\rm se}=a(r)\mathbf
s_{1}\cdot\mathbf s_{2}$, where $a(r)$ is a function of the
internuclear distance \cite{happer_rmp}. For one such collision we denote by $\omega_{\rm se}$
the integral of $a(r)$ over the collision trajectory, hence the
Hamiltonian describing one completed collision is ${\cal H}_{\rm
se}=\omega_{\rm se}\mathbf s_{1}\cdot\mathbf s_{2}$ (in units $\hbar=1$). Obviously $\omega_{\rm se}$
depends on the particular collision trajectory. If $\tau_{c}$
is the duration of the collision, then $\phi_{\rm se}=\omega_{\rm
se}\tau_{c}$ is the phase angle swept by each atomic spin during
this collision. Due to the electrostatic nature of spin-exchange
collisions \cite{happer_rmp}, $\phi_{\rm se}\gg 1$. By measuring $\phi_{\rm se}$ of
atom 2 (the quantum probe), we can in principle extract
information about the spin state of atom 1 (the quantum system).
Indeed, the interaction Hamiltonian ${\cal H}_{\rm se}$ is
interpreted by atom 2 as an effective magnetic field
$\mathbf{B}=\omega_{\rm se}\langle\mathbf{s}_{1}\rangle$, hence
$\phi_{\rm se}$ is the precession angle of atom 2 spin in this
magnetic field. Although extracting the value of $\langle\mathbf{s}_{1}\rangle$ requires
knowledge of the specific collision trajectory (hidden in the
precise value of $\omega_{\rm se}$), the direction of
$\langle\mathbf{s}_{1}\rangle$ can be readily found from the sign
of the phase rotation $\phi_{\rm se}$. Another quantum probe
(another atom) can extract similar directional information at a
later time. A large number of such collisions is thus found to
sample the atomic spin precession, hence from a series of such
observations the spin-precession (Larmor) frequency $\omega$ can be inferred.
Needless to mention that this is not the way that $\omega$ is measured in actual
experiments. However, information being physical \cite{chuang}, the particular way of extracting it is
inconsequential.

The uncertainty in such a measurement of $\omega$ will be
determined by the fact that the measurement cannot go on forever.
Spin-exchange collisions will eventually produce a back-action on the
measured quantum system. For small times and large spin-polarizations, this back-action is minimal \cite{walter}. However, as the spin-polarization decays, spin-exchange collisions will be able to
induce a large phase jump $\phi_{\rm se}$ on the coherent
spin precession of atom 1 (or, equivalently, any other atom).
 The number of such phase jumps per unit
time will be given by the spin-exchange rate $\gamma_{\rm
se}=nv\sigma_{\rm se}$, where $n$ is the atom number density, $v$
the mean relative velocity of two colliding atoms and $\sigma_{\rm
se}$ the spin-exchange cross section. At long times, when the spin-polarization has decayed away, the measurement of the sign of $\langle\mathbf{s}_{1}\rangle$ will merely reflect spontaneous spin noise \cite{katso,qrng}.
This collision-induced sampling process will thus result in a distribution of measured precession frequencies, the
width of which will be on the order of $\gamma_{\rm
se}$. This is the spin-exchange broadening that
limits the precision with which one can measure $\omega$
\cite{note1}. From the view point of quantum measurements
performed on an atom, the spin-exchange rate $\gamma_{\rm
se}$ is identified with the measurement rate, i.e. the rate at
which we extract information about the spin state of any given atom.
An unexpected phenomenon is observed when $\gamma_{\rm se}\gg\omega$:
the width of the spin-resonance shrinks and scales as $\omega^{2}/\gamma_{\rm se}\ll\gamma_{\rm se}$ \cite{happer_tang,happer_tam}.
This is the quantum Zeno effect observed in the strongly interrogated atomic spin coherence. This dependence of the suppressed decoherence rate, i.e. the $\omega^{2}/\gamma$ dependence is exactly the characteristic dependence of quantum Zeno effect, as has been described in \cite{braginsky}
and more recently in \cite{pritch}.
It can be rephrased as follows: if the rate of performing measurements on (or extracting information from) a coherently evolving quantum system is larger than the system's evolution rate, the measurement-induced back-action on the system is suppressed.
\begin{figure}
 \centering
 \includegraphics[width=8 cm]{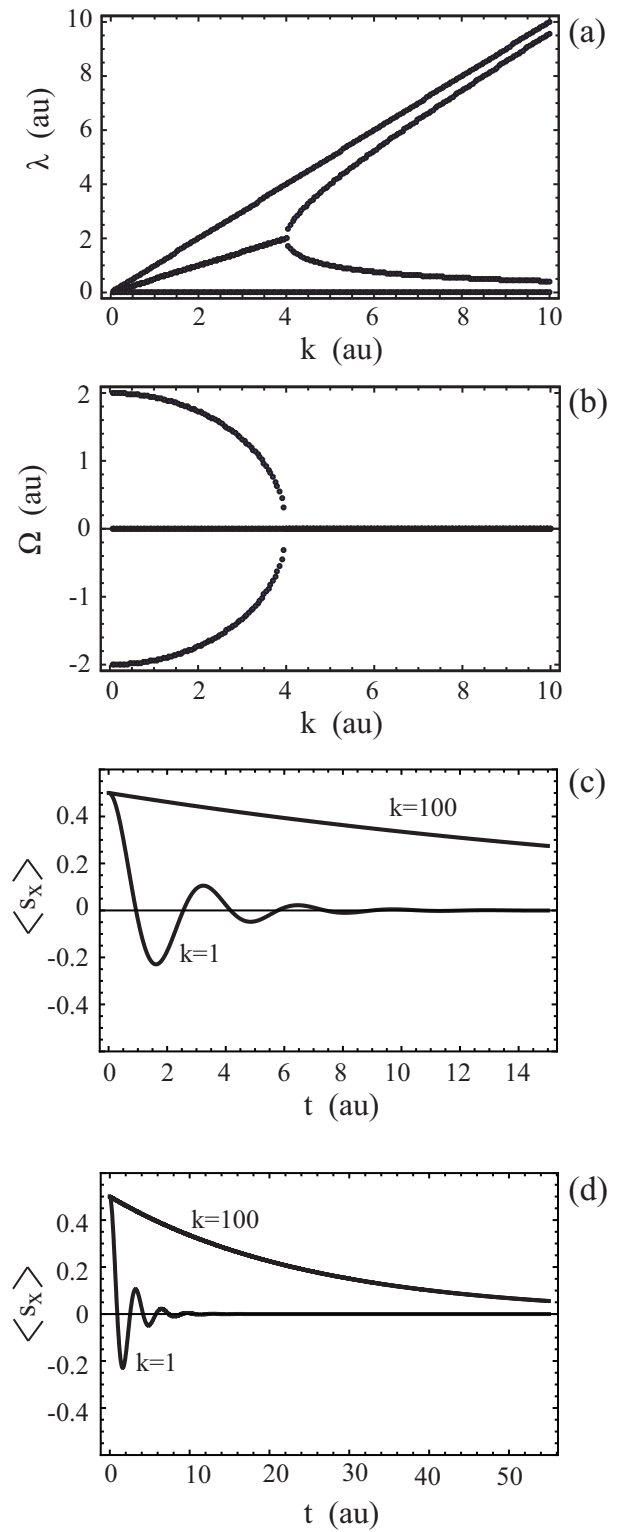}
 \caption{Decay rates (a) and precession frequencies (b) corresponding to the eigenvalues of Eq. (\ref{eq:rho}) for $\omega=1$. (c) Time evolution of the
 expectation value $\langle s_{x}\rangle$ for $\omega=2$ and two different values of the measurement rate $k$. The linear early-time dependence for $k=100$ is evident. (d) same as before, but for longer times.}
 \label{fig1}
\end{figure}
\section{Quantitative Arguments}
Towards a simplified quantitative argument, we describe the effects of collision-induced measurements on the atomic spin state by the density matrix equation
\beq
d\rho/dt=-i[H,\rho]-k[s_{x},[s_{x},\rho]]\label{eq:rho}
\eeq
where $H=\omega s_{z}$ is the Zeeman interaction Hamiltonian, and the second (dissipative) term takes into account \cite{braginsky} the measurement of $s_{x}$ at a rate $k$. In Fig. \ref{fig1}(a) and Fig. \ref{fig1}(b) we show the decay rates $\lambda$ and precession frequencies $\Omega$ of the
four complex eigenvalues of (\ref{eq:rho}), which are of the form $-\lambda+i\Omega$. The calculation was performed for constant $\omega=1$. It is evident that one of the decay rates is suppressed when the measurement rate $k\gg\omega$. In Fig. \ref{fig1}(c,d) we show the time dependence of the
expectation value $\langle s_{x}\rangle$, for two different values of the measurement rate $k$. It is seen that while at small values of
$k$ the coherent precession of $\langle s_{x}\rangle$ decays at a rate proportional to $k$, at high measurement rates $\langle s_{x}\rangle$ survives for a much longer time (in this simple model the precession frequency $\Omega$ is also suppressed). In Fig. \ref{fig1}(c) in particular, the initial linear decay (in the case $k=100$) is clearly seen.

In reality, the effect of spin-exchange is described by a non-linear density matrix equation, that leads to similarly suppressed decay rates \cite{happer_tam}. Moreover, spin-exchange collisions are different from the kind of measurements usually considered \cite{itano} in that they do not collapse the wavefunction to the initial state, but make atoms quantum-jump from one ground-state hyperfine-multiplet to the other. The Larmor spin precession has opposite sense in the these two
multiplets. However, the analog of the probability to find the system in the initial state which is usually considered in quantum Zeno effects \cite{itano}
is in this case found in the correlation of the spin-coherence, i.e. the overlap between an unperturbed spin precession and one including such collision-induced jumps. Specifically, if we write $\sigma(t)=\cos\omega t$ for the expectation value of the Pauli operator $\sigma_{x}$, and $\sigma'(t)$ is the same function but including the occurrence of a jump in the precession frequency from $\omega$ to $-\omega$ at time $\tau$, then the average value of the correlation $p=(1/2\tau)\int_{0}^{2\tau}{\sigma(t)\sigma'(t)dt}$ can be approximated for $\omega\tau\ll 1$ by
\beq
p\approx 1-(\frac{\omega\tau}{2})^{2}
\eeq
After $N$ such independent collisions taking place in a total time interval $T=N\tau$, the overlap between the initial unperturbed precession and the one including $N$ collisions will have
decayed to
\beq
P=\left[1-(\frac{\omega\tau}{2})^{2}\right]^{N}\approx e^{-(\omega^{2}\tau/4)T}
\eeq
Thus we recover the decay rate $\omega^{2}/4\gamma_{\rm se}$, where $\gamma_{\rm se}=1/\tau$ is the spin-exchange rate. This rather simplified analog of the rigorous statistical treatment \cite{happer_tam} of spin-exchange collisions is meant to point out the dependence $\omega^{2}/\gamma$
common to all appearances of the quantum Zeno effect in quantum systems characterized by an intrinsic frequency scale $\omega$ and a measurement rate $\gamma$.
\newline
\section{Absence of information in spin-destruction collisions}
Contrary to spin-exchange collisions which dissipate only spin coherence, there is another kind of binary collisions relaxing populations as well as coherences: the spin-destruction collisions \cite{spin_axis}, described by an interaction of the form
${\cal H}_{\rm sd}=6\lambda(r)(\mathbf{s}_{1}\cdot\mathbf{\hat{r}})(\mathbf{s}_{2}\cdot\mathbf{\hat{r}})-2\lambda(r)\mathbf{s}_{1}\cdot\mathbf{s}_{2}$, where $\mathbf{\hat{r}}$ is the unit vector along the internuclear axis, and $\lambda(r)$ is a function of the internuclear distance. In the first term of ${\cal H}_{\rm sd}$ we have the direct participation of the environment degrees of freedom,
i.e. it is this term that opens the loss-channel of spin-angular momentum into translational angular momentum. It is clear that not even the sign of $\langle\mathbf{s}_{1}\rangle$ can be inferred by observing the phase jump of atom 2 spin, since the effective magnetic field seen by atom 2 is now proportional to $6(\langle\mathbf{s}_{1}\rangle\cdot\mathbf{\hat{r}})\mathbf{\hat{r}}-2\langle\mathbf{s}_{1}\rangle$. Since these two terms are of similar magnitude, the information loss into the environment is dominant no matter what the collision rate is.


\begin{thebibliography}{9}
\bibitem{note1}
This broadening is accompanied by the so-called spin-exchange
frequency shift, that limits the accuracy of atomic clocks.
However this effect is not relevant for the present discussion. For more details see
Ref. \cite{walter}.
\end{thebibliography}
\end{document}